\documentclass[pre,aps,preprint,showpacs]{revtex4}
\usepackage{revsymb,amsmath}
\usepackage{graphicx}
\usepackage{bbm}
\begin{document}
\author{Michael Schulz }
\affiliation{University Ulm\\D-89069 Ulm Germany}
\email{michael.schulz@uni-ulm.de}
\author{Steffen Trimper}
\affiliation{Institute of Physics,
Martin-Luther-University, D-06099 Halle Germany}
\email{steffen.trimper@physik.uni-halle.de}
\title{Exact expression for Drude conductivity in one-dimension with an arbitrary potential}
\date{\today }

\begin{abstract}
\noindent An exact expression for the Drude conductivity in one dimension is derived under the 
presence of an arbitrary potential. In getting the conductivity the influence of the electric field 
on the crystal potential is taken into account. This coupling leads to a systematic deformation 
of the potential and consequently to a significant modification of the charge transport. The corrections to the 
conventional Drude conductivity are determined by the configurational part of the partition function. The 
activation energy for the conductivity process is expressed by a combination of the free energy of the 
underlying equilibrium system. The electric current is calculated in the linear response regime by solving the 
Smoluchowski equation. The steady state solution differs significantly from the equilibrium distribution.
In case of a tight binding potential the conductivity offers corrections depending on the amplitude of 
the potential. As a further application we discuss nanocontacts with piecewise constant potentials. 
The electric conductivity is corrected by the potential height. 
\end{abstract}


\maketitle

\section{Introduction}
\noindent The Drude model of electric conductivity \cite{d} offers a rather generic behavior for 
charge transport. Although the concept of the Drude conductivity seems to be old-fashioned it is 
very useful for different applications, so for chaotic transport and weak localization \cite{jw}, 
magnetoresistance \cite{cdk}, dynamical ordering in a confined Wigner crystal \cite{gd} and 
optical conductivity \cite{s}. More recently the Drude model came into the focus in describing 
the spin Hall effect \cite{c}, for a more detailed consideration see \cite{erh}. In spite of being 
a classical concept the Drude model yields the universal dc and ac conductivity and its temperature 
dependence \cite{ma}. In particular, the conductivity is expressed by the charge $e$, the mass $m$ 
and the density $n$ of the charge carriers as well as a relaxation time $\tau$ via $\sigma_D = n e^2 \tau/m$. 
In general the parameter $\tau$ reflects the microscopic origin of the transport process such as 
different scattering mechanisms. In the Markov approach the relaxation time is given in terms of 
a simple damping constant $\gamma = \tau^{-1}$. However, the Drude conductivity should be influenced 
and corrected by the underlying periodic crystal potential. And even that point we would like to discuss 
in the present paper. Here the Drude conductivity is generalized by including the lattice potential. 
The analysis is based on the Smoluchowski equation \cite{gar} which yields likewise the probability density 
for the electrons in the lattice. As the result we get an exact expression for the charge conductivity 
$\sigma = j/E$ in the limit $E \to 0$. The final relation is illustrated for realistic model potentials. 
Especially, the approach enables us to study the conductivity of nanocontacts by assuming piecewise constant 
potentials. However to find out the conductivity one is confronted with the problem, that the applied external 
electric field $E$ exhibits a feedback to the potential. The potential is deformed under the influence of 
the field which leads to a modification of the motion of the charge carriers. Different to the quantum 
approach the carriers can overcome barriers or traps not by tunneling processes but by thermal activation.     

\section{Model}

\noindent The classical equation of motion for the charge carriers under consideration of thermodynamic 
fluctuations are 
\begin{equation}
\gamma \dot{x}(t) =  \frac{\partial \mathcal{H}}{\partial x} + \eta(t)\quad\rm{with}\quad
\langle \eta (t) \eta (t') \rangle = 2\, D\, \delta(t - t')\,. 
\label{kan}
\end{equation}
Here $\gamma$ is a damping parameter and $D$ is the strength of the Gaussian process. Both of them model the interaction with 
a heat bath. The Hamiltonian $\mathcal{H}$ is given by  
\begin{equation}
\mathcal{H} = \frac{m}{2} v^2 + U(x) - e E x\,,
\label{ham}
\end{equation}
where $E$ is a homogeneous electric field and $U(x)$ is the periodic crystal potential which is originated by 
all other particles of the system, but one can also include other potentials such as defect potentials. 
Although in the Hamiltonian the contribution of the potential and the electric field energy are well 
separated the potential may modify by the field as demonstrated. The Smoluchowski equation equivalent 
to eq.~\eqref{kan} reads \cite{gar} 
\begin{equation}
\frac{\partial \rho (x,t)}{\partial t} = - \frac{\partial j_p(x,t)}{\partial} \quad\rm{with}\quad
j_p(x,t) = \left\{ - D \frac{\partial }{\partial x} - \frac{1}{m \gamma} 
\left[ \frac{\partial \mathcal{H}}{\partial x}  \right] \right\} \rho (x,t)\,.
\label{fp}
\end{equation}
Inserting the Hamiltonian eq.\eqref{ham} the probability current $j_p$ can be written as 
\begin{equation}
j_p(x,t) = - D \frac{\partial \rho(x,t)}{\partial x} - \frac{1}{m \gamma} 
\left [ \frac{\partial U(x)}{\partial x} - e E   \right] \rho(x,t)\,.
\label{strom}
\end{equation}
Here $\rho(x,t)$ is the probability density 
In case the system is coupled to a heat bath with temperature $T$ the equilibrium solution 
is given by the 
\begin{equation}
\rho_e(x) = \mathcal{N} \exp( - \beta \mathcal{H})\quad \beta^{-1} = k_B T\,,
\label{eq}
\end{equation}
provided the Einstein relation $D = k_B T/\gamma m $ is fulfilled. $\mathcal{N}$ is a 
normalization constant. Notice that the equilibrium distribution satisfies $j_p = 0$ whereas a 
steady state solution $\rho_s(x)$ obeys the weaker condition $\partial_x j_p = 0$. It is easy to 
see that the equilibrium solution $\rho_e$ has a physical meaning only in case of a vanishing 
electric field. Otherwise, the probability density $\rho_e$ is due to the presence of the electric field 
not normalizable. If there exist an equilibrium state for nonzero field all charged particles would 
accumulate at $x \to - \infty$. It should be remarked that the same situation occurs in a quantum treatment 
of the problem where one can not find a wave function normalized 
in the whole space. However, different to the quantum case the statistical approach allows a steady state 
solution describing the physical situation adequately. The electric current is given by 
\begin{equation}
j_{el}  =  n e \langle v \rangle = \frac{N e}{L} \int_0^L dx j_s(x)\,.
\label{dru}
\end{equation}
Here, $N$ is the total number of charge carriers and $n$ is the corresponding density. 
From here we conclude immediately that the equilibrium distribution \eqref{eq} is an improper distribution 
function because the current is zero due to the symmetry $\mathcal{H}(v) = \mathcal{H}(-v)$.

\section{Electrical Conductivity}

\noindent In order to obtain the electrical conductivity one has to apply the steady state solution of 
Eq.~\eqref{fp}. For a zero potential $U = 0$ we find the constant steady state current by integration of 
Eq.~\eqref{strom}
\begin{equation}
j_{el} = \frac{N e}{L}\left[ D (\rho(0) - \rho(L)) + \frac{e E }{\gamma m} 
\int_0^L \rho(x) dx \right]\,. 
\end{equation}
$N$ is the total number of charge carriers and $L$ is the size of the system. Taking into account 
periodic boundary conditions and the normalization condition for $\rho(x)$ one obtains
\begin{equation}
j_{el} =  \sigma_D E\quad{\rm with}\quad \sigma_D = \frac{ n e^2 }{m \gamma}\,.
\label{dru1}
\end{equation}
The quantity $n = N/L$ is the particle density. The last relation is nothing else than the Drude conductivity 
where the relaxation time is given by the inverse damping parameter 
$\gamma$. For a non-zero potential the relation in Eq.~\eqref{dru1} will be modified accordingly. 
Directly by integration it follows  
\begin{equation}
j_{el} = \sigma_D E - \frac{N e}{m \gamma L} \int_0^L dx \frac{dU(x)}{dx} \rho_s(x) = 
\sigma_D E - \frac{N e}{m \gamma L} \left \langle \frac{d U}{dx} \right \rangle \,. 
\label{dru2}
\end{equation}
The average is taken using the field dependent steady state probability distribution function 
$\rho_s(x) \equiv \rho_s(x; E)$. To find out the contribution induced by the electric field we need the 
steady state solution for the current which obeys $\partial_x j_p(x) \neq 0$. 
Because the steady state solution $\rho_s$ depends likewise on the electric field we have to incorporate 
the influence of the electric field on the averaged crystal potential, compare Eq.~\eqref{dru2}. 
Since we are interested in the conductivity we need the steady state solution for a weak electric field $E$ 
only. Therefore we make the ansatz 
\begin{equation}
\rho_s(x; E) = \mathcal{N}(E) e^{-\frac{U(x)}{k_BT}} [ 1 + E \phi(x) ]\,,
\label{stea}
\end{equation}  
with an arbitrary function $\phi(x)$ and a normalization factor $\mathcal{N}(E)$. Inserting Eq.~\eqref{stea} in 
Eq.~\eqref{strom} we find 
\begin{equation}
\frac{d \phi(x)}{dx} = \frac{e}{k_B T} - \frac{j_p }{ E D \mathcal{N}(E)} \exp(U(x)/k_B T)\,. 
\label{stea1}
\end{equation}
The current $j_p$, defined in Eq.~\eqref{strom}, disappears in for zero field. Therefore we write for small 
electric field $j = a E$ with the unknown parameter $a$ which is determined below. The solution of the 
Eq.~\eqref{stea1} is
\begin{equation}
\phi(x) = \phi_0 + \frac{e x}{k_BT} - \frac{a}{D \mathcal{N}(E)} \int_0^x dx' \exp(U(x')/k_BT)\,.
\label{1d}
\end{equation}
Imposing periodic boundary conditions it follows
\begin{equation}
\frac{e L}{k_B T} = \frac{a}{D \mathcal{N}(E) } \int_0^L dx \exp(U(x)/k_BT)\,.
\label{stea3}
\end{equation} 
According to Eq.~\eqref{dru2} and Eq.~\eqref{stea} as well as periodic boundary conditions the contribution 
to the conductivity is    
$$
\left \langle \frac{d U}{dx} \right \rangle = \int_0^L dx \frac{d U}{dx} \rho_s(x; E) = k_B T 
\mathcal{N}(E) E \int_0^L e^{-\frac{U(x)}{k_BT}} \frac{d \phi(x)}{dx} dx \,.
$$
Using Eqs.~\eqref{stea}, \eqref{stea1} and \eqref{stea3} one finds 
\begin{equation}
\left \langle \frac{d U}{dx} \right \rangle = k_B T \mathcal{N}(E) E \left[ \frac{e}{k_B T \mathcal{N}(0)}  - 
\frac{a L}{D \mathcal{N}(E)} \right] =  e E \left[ 1 - \frac{L^2}{Z_L[U]\,Z_L[-U]} \right] + O(E^2)
\label{cor}
\end{equation}
with 
\begin{equation}
Z_L[U] = \int_0^L dx \exp( U (x)/k_BT)\,.
\end{equation}
Inserting the result in Eq.~\eqref{dru2} our final relation for the conductivity reads
\begin{equation}
\sigma = \sigma_D \frac{L^2}{Z_L[U]\,Z_L[-U]} \equiv \sigma_D \frac{Z^2[U=0]}{Z_L[U] Z_L[-U]}\,.
\label{fin}
\end{equation}
Because the last relation includes the ratio of the partition functions one can also insert the total 
partition function of the system without the electric field part.
\begin{equation}
Z_L[U] = {\rm Tr} \exp [ - \frac{m v^2}{2} + U(x) ]\,.
\label{fin1}
\end{equation}
The final relation offers the duality property namely a symmetry against $U(x) \to -\, U(x)$. Thus, in our 
approach one can not distinguish the barrier and the trapping problem. Eq.~\eqref{fin} can be also rewritten 
in the form
\begin{equation}
\sigma = \sigma_D e^{- E_A/k_BT}\quad{\rm with}\quad E_A = F[U] + F[-U] - 2 F[U=0],\quad F = - k_B T \ln Z\,.
\label{fin2}
\end{equation}
A typical non-equilibrium quantity as the activation energy $E_A$ is completely expressed by the equilibrium 
free energy $F$. Let us note that the Einstein relation is fulfilled by the equilibrium distribution when 
the system is coupled to a heat bath. Otherwise in the present approach we apply the steady state solution. 
In that case we can set $D \gamma m = \epsilon_0\,$, where $\epsilon_0$ is a characteristic energy of the 
system, for instance the ground state energy of a quantum model.

\section{Model Potentials}

\begin{figure}[h]
\centering
\includegraphics[width=1.0\textwidth]{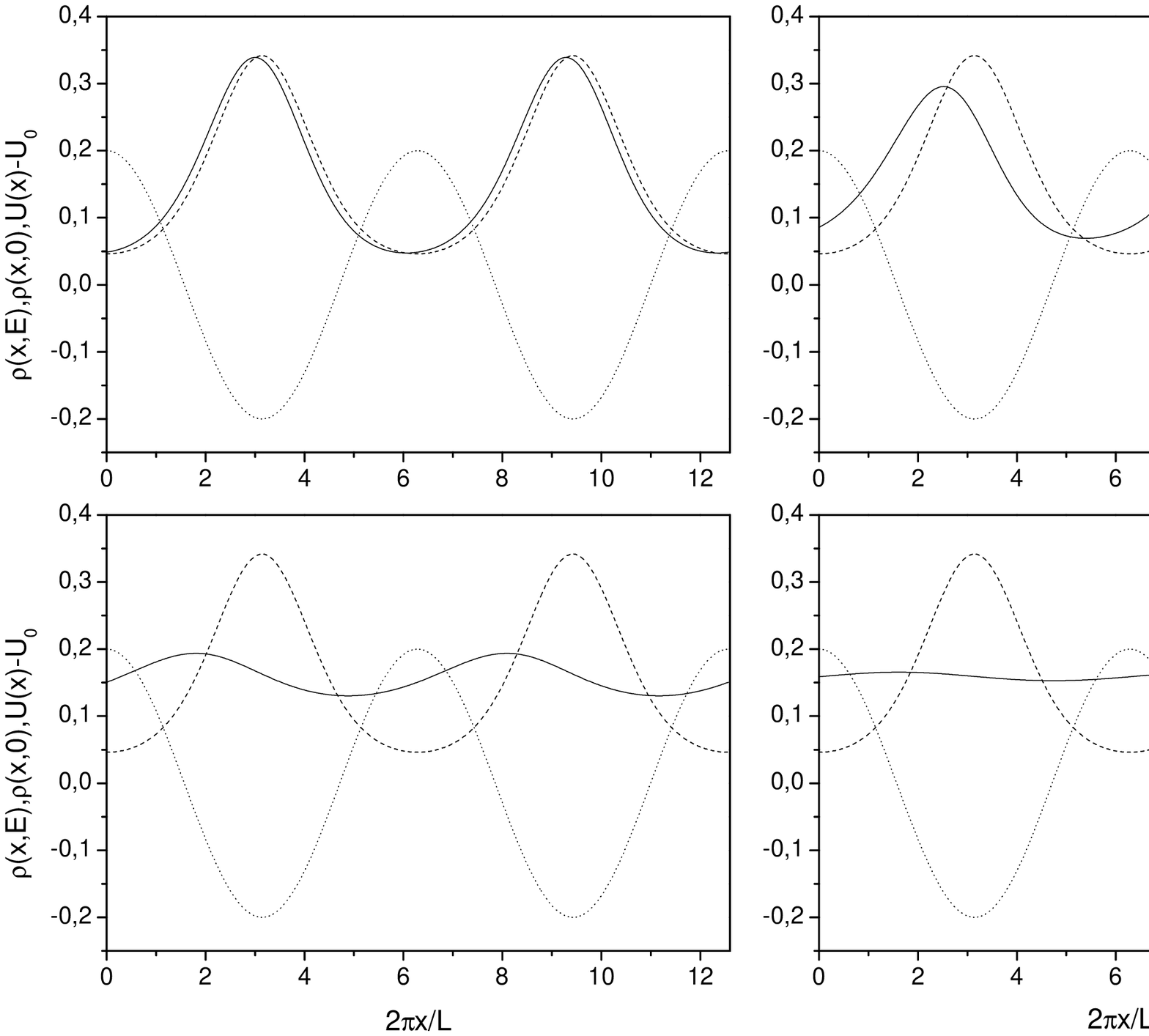}
\vspace{-5cm}\caption{Periodic potential $U(x)$ (dotted line) and probability distribution of the position 
$\rho (x)$ for zero field (dashed lines) and with field (full line). The parameter from left top to right bottom 
is the ratio $\frac{E L}{U_0} = 0.2;\,1.0;\,5.0;\, 25.0$; the position $x$ is given in units of $L/2\pi$\,. }
\label{fig1}
\end{figure}

\begin{figure}[h]
\centering
\includegraphics[width=0.9\textwidth]{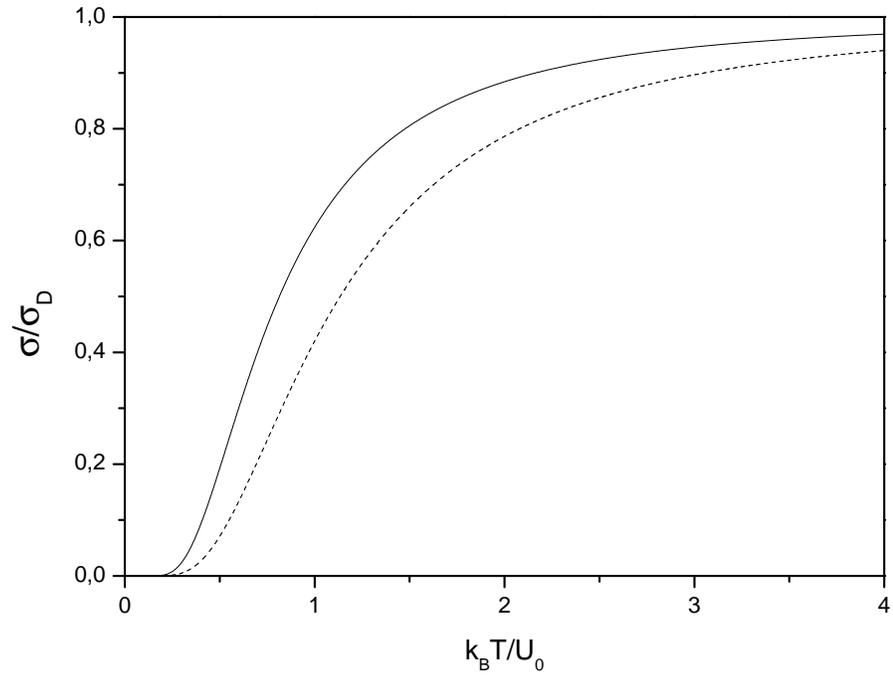}
\vspace{-3cm}\caption{Electric conductivity $\sigma$ for the $\cos$-potential (full line) and the 
piecewise constant potential 
(dashed line)\,. }
\label{fig2}
\end{figure}

\begin{figure}[h]
\centering
\includegraphics[width=0.9\textwidth]{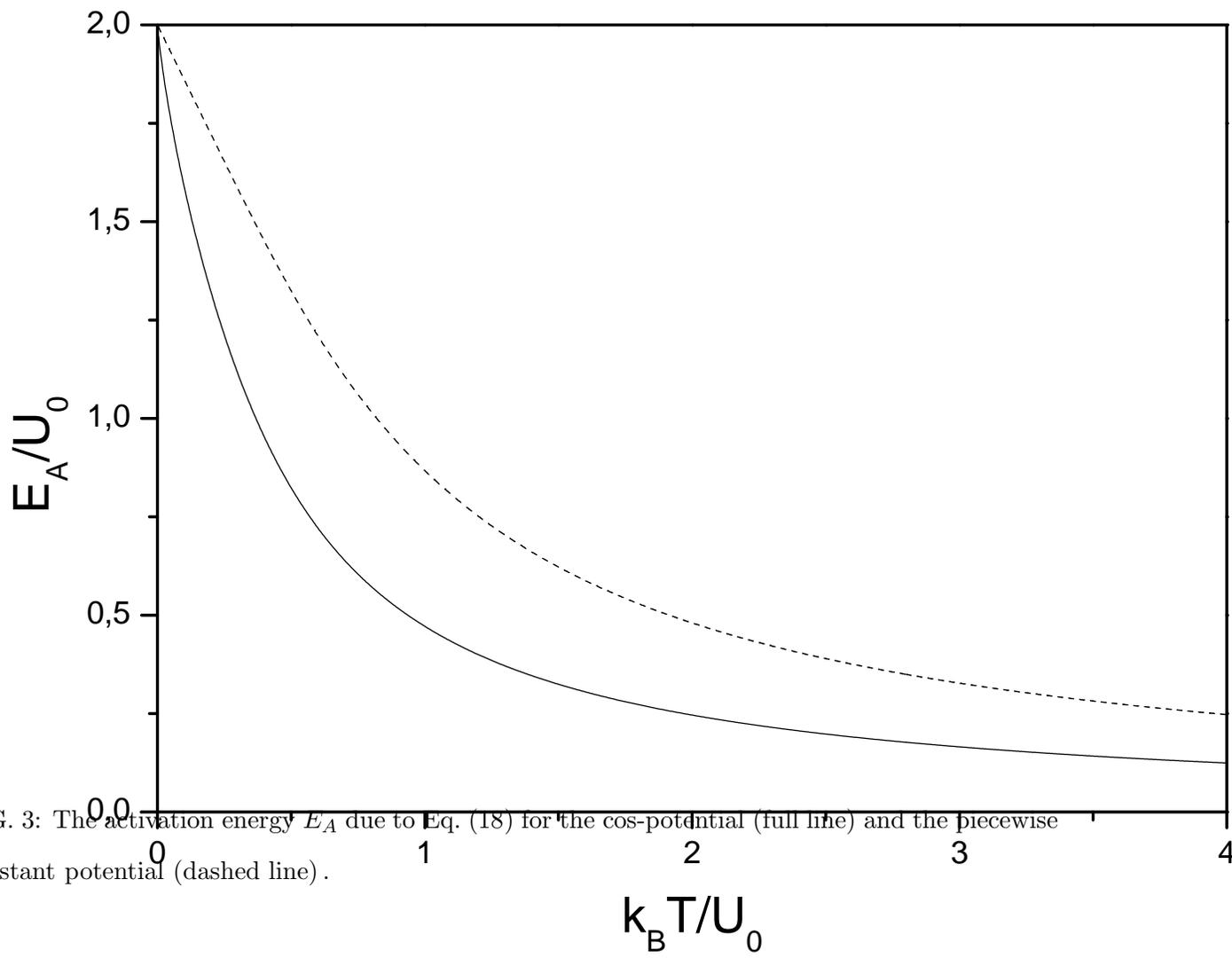}
\vspace{-3cm}\caption{The activation energy $E_A$ due to Eq.~\eqref{fin2} for the $\cos$-potential (full line) and 
the piecewise constant potential (dashed line)\,. }
\label{fig3}
\end{figure}

\noindent Now let us illustrate our approach in two examples. Firstly we 
consider the tight binding potential, dotted line in Fig.~\ref{fig1}:
$$
U(x) =U_0\,[ 1 - \cos(qx)\,]\quad q=\frac{2\pi}{L}\,.
$$
The electric field causes a systematic shift of the potential due to Eq.~\eqref{cor}. As the result the 
change of the probability density for the position $\rho_s(x;E)$ according to Eq.~\eqref{stea} is shown 
in Fig.~\ref{fig1}. With increasing strength of the electric field $E$ the shift becomes more pronounced. 
The shift of the potential or the probability, respectively is so organized that a constant current is 
maintained. For a small field in terms of $U_0/L$ the charge carriers follow immediately the potential. 
The largest probability to find the electron is at the minimum of the potential. When the field strength 
is enlarged then the probability density for the position is shifted considerably. For very high field the 
charge carriers are actually not influenced by the potential. In that case the probability density is 
nearly constant as shown in the last graph in Fig.~\ref{fig1}. The analytical calculation for the 
conductivity yields according to Eq.~\eqref{fin}   
\begin{equation}
\sigma = \frac{\sigma_D}{I_0^2(\frac{U_0}{k_BT})} \,.
\label{tb}
\end{equation}
Here $I_0(y)$ is the Bessel function. In the low temperature case $U_0 \gg k_B T$ the conductivity behaves 
as $\sigma \simeq \sigma_D \exp( - 2 U_0/k_BT)$, i. e. the activation energy due to Eq.~\eqref{fin2} is 
dominated by the amplitude of the periodic potential $ E_A \simeq 2 U_0 $. In the opposite case of 
high temperatures it results $ E_A \simeq U_0^2/2 k_B T $. The conductivity offers a small 
correction $ \sigma \simeq \sigma_D [ 1 - (U_0/ k_BT)^2/2 \,]$. The conductivity is depicted 
in Fig.~\ref{fig2} and the activation energy in Fig.~\ref{fig3}.\\
A second illustration is given by a sequence of $N$ nanocontacts modeled by a piecewise constant potential,
$U(x) = U_0\quad 0\leq x \leq a$ and $U = -U_0 \quad a \leq x \leq 2a$ and periodic continuation with $L = N a$. 
The number of barriers and the number of traps are both $N/2$.  The conductivity follows from Eq.~\eqref{fin} to 
\begin{equation}
\sigma = \sigma_D \frac{2}{\,1 + \cosh(\frac{2 U_0}{k_B T})}\,.
\label{nano}
\end{equation}
The activation energy can be expressed by 
\begin{equation}
E_A = k_B T \ln\left( \frac{1 + \cosh(2 U_0/k_B T)}{2} \right)
\end{equation}
The limiting cases are
$$
\frac{\sigma}{\sigma_D} \simeq 4\,e^{- 2 U_0/k_B T}\quad{\rm if}\quad U_0 \gg k_B T\,;\quad 
\frac{\sigma}{\sigma_D} \simeq 1 - \left(\frac{U_0}{k_B T}\right)^2 \quad{\rm if}\quad U_0 \ll k_B T\,.
$$
For an infinite high barrier the material becomes an insulator, whereas for zero potential 
the conventional Drude conductivity appears. The conductivity for the piecewise 
constant potential is shown in Fig.~\ref{fig2} as the dashed line. 
The conductivity increases continuously with increasing temperature. The higher the temperature in comparison 
to the potential height $U_0$ the more charge carriers are able to overcome the barrier. For rather high 
temperature the conductivity becomes constant, i. e. the influence of the potential is negligible. 
The same situation is observed for the $\cos$-potential (full line in Fig.~\ref{fig2}. The increase of 
the conductivity is accompanied with a decrease of the activation energy $E_A$. The behavior of $E_A$ for both 
the piecewise constant potential (dashed line) and the $\cos$-potential is shown in Fig.~\ref{fig3}. 
For practical purposes let us consider the case that there exists only one barrier of height $U_B$ and width 
$\triangle $. The length of each of the two input leads is assumed to be $l$ and the corresponding heights are 
$U_0$. Introducing the dimensionless ratio $\kappa = \triangle /2 l$ 
the conductivity can be written as 
$$
\sigma = \sigma_D \frac{(1 + \kappa )^2}{\,(1 - \kappa )^2 + 4 \kappa \frac{\sigma_D}{\sigma_m}}  
$$  
where $\sigma_m$ is the minimal conductivity realized for $\kappa = 1 $. It holds 
$$
\sigma_m = \frac{ 2 \sigma_D}{ 1 + \cosh(U_B - U_0)/k_B T)} 
$$
For $\kappa \ll 1$ one gets
$$
\frac{\sigma}{\sigma_D} \simeq 1 - 4 \kappa \left[ \frac{\sigma_D}{\sigma_m} - 1 \right]\,.
$$
In the same manner one can find the conductivity for any other potential. 

\section{Conclusion}

\noindent In the present paper a simple classical model for the electrical conductivity is  
proposed and solved in one dimensions. Regarding nanocontacts and nanowires such one dimensional  
models are in the focus of recent interests. Here we get an exact expression for the conductivity 
where systematic contributions originated by the underlying periodic potential are taken into account.
To be specific the potential is subjected to a systematic alteration caused by the applied external field. 
The probability distribution for the position of the charged particles is modified by the field. 
Simultaneously the charge carriers are subjected to an effective averaged potential which is likewise 
modified by the field. The alteration is so organized that a constant current is maintained. 
The contribution of the crystal potential is significant for low temperatures whereas in the 
high temperature limit the influence of the potential is weak and the conventional Drude 
conductivity is recovered. Although our approach is based on a classical stochastic one the 
conductivity can be also calculated for a piecewise constant potential which models nanoconducts or 
nanowires. The present analysis is restricted to the one-dimensional case. The three dimensional case is more 
complicated because the steady state solution has to satisfy the condition $\nabla \cdot \vec j_p = 0$ which is 
fulfilled by $\vec j_p = \nabla \times \vec A$ with an arbitrary vector field $\vec A$. In a forthcoming paper 
\cite{st} we study this problem in view of the spin Hall effect applying a similar approach. We are able to estimate 
the contributions of the potential to the conductivity.\\ 
Let us finally remark that a more general expression for the conductivity is obtained 
when the Einstein relation $D \gamma m = k_B T$ is not longer fulfilled. Then for example the conductivity 
for the piecewise constant potential reads
$$
\sigma = \sigma_D \frac{2}{\,1 + \cosh(\frac{2 U_0}{\epsilon_0})}\,,
$$
whereas the characteristic energy scale $\epsilon_0$ can not be determined within our approach. However 
it is appropriate to identify $\epsilon_0$ with the ground state energy in presence of the potential.\\  

\noindent The work has been supported by the DFG: SFB 418

\newpage
~\\~\\~\\~\\~\\~\\ 
\clearpage
~\\~\\~\\~\\~\\~\\
\clearpage
~\\~\\~\\~\\~\\~\\
\end{document}